\documentclass[12pt]{article}

\usepackage{amssymb}
\usepackage{amsmath}
\usepackage{latexsym}

\catcode`@=11
\renewcommand{\section}{\@startsection{section}{1}{0pt}{\medskipamount}
{\medskipamount}{\large\bf}}
\numberwithin{equation}{section}
\catcode`@=12

\def\beq{\begin{eqnarray}}    
\def\eeq{\end{eqnarray}}      

\def\ln{\,\mbox{ln}\,}                  
\def\sTr{\,\mbox{sTr}\,}                

\def\div{\,\mbox{div}\,}                
\def\pa{\partial}                       

\def\={\ =\ }

\begin{document}

\begin{center}

{\Large\bf Does the nontrivially deformed field-antifield
formalism exist?}

\vspace{18mm}

{\Large Igor A. Batalin$^{(a,b)}\footnote{E-mail: batalin@lpi.ru}$\;
and Peter M. Lavrov$^{(b, c)}\footnote{E-mail:
lavrov@tspu.edu.ru}$\; }

\vspace{8mm}

\noindent ${{}^{(a)}}$
{\em P.N. Lebedev Physical Institute,\\
Leninsky Prospect \ 53, 119 991 Moscow, Russia}

\noindent  ${{}^{(b)}}
${\em
Tomsk State Pedagogical University,\\
Kievskaya St.\ 60, 634061 Tomsk, Russia}

\noindent  ${{}^{(c)}}
${\em
National Research Tomsk State  University,\\
Lenin Av.\ 36, 634050 Tomsk, Russia}

\vspace{20mm}

\begin{abstract}
\noindent
We reformulate the Lagrange deformed field-antifield BV -formalism
suggested,  in terms of
the general Euler vector field $N$ generated by the antisymplectic potential.
That $N$ generalizes,
in a natural anticanonically-invariant manner,  the usual power-counting
operator. We provide
for the "usual" gauge-fixing mechanism as applied to the deformed BV
-formalism.
\end{abstract}

\end{center}

\vfill

\\


\section{Introduction}

In the field-antifield formalism \cite{BV,BV1,BV2}, the concept of
deformations based on a nilpotent higher-order operator $\Delta$ was
developed in a series of articles
\cite{BT3,Ak,AD,BDA,BBD1,BBD3,BM1,BM2,BM3}. Such deformations
usually modify the Jacobi identity with BRST exact terms. In
contrast to that, one can, with no assumptions {\it a priori} on
underlying $\Delta$ operator, consider "local" deformations of the
antibracket with a Boson deformation parameter, such that the Jacobi
identity holds strongly.

Historically,  the deformed antibracket has been studied in the
articles \cite{LSh,KoT,KoT1,KoT2}. In Ref. \cite{BB2}, the deformed
$\Delta$-operator has been found that differentiates the deformed
antibracket, and the first attempt has been made to understand
actually possible role of the deformed antibracket and
$\Delta$-operator  in the construction of the $W-X$ version
\cite{BT1,BT2,BT3,BMS,BT4,BBD1,BBD2,BB3} of the  Lagrange deformed
field-antifield Batalin-Vilkovisky (BV)-formalism \cite{BB2}. If one
believes that the deformed BV - formalism still describes
gauge-invariant field systems, then there appears a difficult
problem of how the deformation can coexist with the usual
gauge-fixing mechanism. Or, in other words, if that is possible to
provide for a  proper solution to the deformed classical/quantum
master equation.  In the present article, we will  try to pay
attention enough to seek for a possible way to resolve the mentioned
problem.

In principle, our present consideration is based essentially on the logic
and mathematics of the article \cite{BB2} of Batalin and Bering.
Regrettably, these authors did not
accomplish their task of construction of the nontrivially deformed field-antifield
formalism
based on the deformed antibracket and $\Delta$ operator. The main idea was to extend the
original antisymplectic phase space with a single extra field-antifield pair just
controlling the scale of deformation. Then, one defines a trivial deformation
within the extended phase space,
and then one should reduce effectively the scale of trivial deformation in
such a way that the latter becomes nontrivial in a consistent manner. That idea seems
promising, the same as before. However, there remains a difficult unresolved problem of
consistent coexistence between the  properness principle and non-triviality of
the deformation. We would like to try again to attack that problem.

\section{Extended $\Delta$-Operator}

We begin with the standard odd Laplacian operator,
\beq \label{i2}
\Delta = \frac{1}{2}(-1)^{\varepsilon_{A}} \pa_{A} E^{AB} \pa_{B},\quad
\pa_A= \frac{\pa}{\pa Z^A}, \quad
\varepsilon(\Delta) = 1,  \quad \Delta^{2} = 0,   
\eeq where $Z^{A}$ , $\varepsilon_{A} = \varepsilon( Z^{A})$, are
original  Darboux coordinates of the field-antifield formalism, and
$E^{AB}$ is a constant invertible antisymplectic metric with the
usual  statistics and dual antisymmetry properties,
\beq
\label{2.2}
\varepsilon( E^{AB} ) = \varepsilon_{A} + \varepsilon_{B} + 1, \quad
E^{AB} = - E^{BA}(-1)^{ (\varepsilon_{A} + 1) (\varepsilon _{B} +
1)}.
\eeq
In what follows below, typical functions, depending only
on the $Z^{A}$ variables of the original sector, will be denoted in
small letters. Now, let us extend the original $Z$-sector by
including two new variables, a Boson $t$ and a Fermion $\theta$, to
extend the original odd Laplacian, to become \cite{BB2} (see also
Appendix A)
\beq
\label{i1} \Delta_{\tau} = t^{2} \Delta + N_{\tau}
\pa_{\theta}, \quad
 \pa_{\theta}=\frac{\pa}{\pa \theta}, \quad
 \varepsilon(\Delta_{\tau}) = 1,\quad \Delta^2_{\tau} = 0, 
\eeq
where
\beq
\label{i3}
N_{\tau} = N + t \pa_{t}, \quad N = N^{A} \pa_{A}.       
\eeq
In what follows below, typical functions depending on the full set of variables,
$Z^{A}, t, \theta$, will be denoted in capital letters.
Nilpotency of $\Delta_{\tau}$ requires
\beq
\label{i4}
[ \Delta, N ] = 2 \Delta,       
\eeq
or, in more detail,
\beq
\nonumber
&&[ \Delta, N ] = [ \Delta, N^{A} ] \pa_{A} = \left[ ( \Delta N^{A} ) +
{\rm ad}( N^{A} ) (-1)^{\varepsilon_{A}} \right]\pa_{A} = \\
\nonumber
&&=( \Delta N^{A} ) \pa_{A} +
\frac{1}{2} \left[ ( N^{A}, Z^{B} ) -
( A \leftrightarrow B ) (-1)^{ (\varepsilon_{A} +1) (\varepsilon_{B} +1) } \right]
(-1)^{\varepsilon_{A}} \pa_{B}\pa_{A} =\\
\label{i5}
&&=2\Delta  =  E^{AB} (-1)^{\varepsilon_{A}} \pa_{B} \pa_{A},      
\eeq
which, in turn, implies
\beq
\label{i6}
\Delta N^{A} = 0,         
\eeq
and
\beq
\label{i7}
E^{AB} =\frac{1}{2} \left[ ( N^{A}, Z^{B} ) - ( A \leftrightarrow B )
 (-1)^{ ( \varepsilon_{A} +1 ) ( \varepsilon_{B} + 1 ) } \right].     
\eeq
Here and below the notation
\beq
\label{ad}
{\rm ad}(F)(...)=(F,(...))
\eeq
for the left adjoint of the antibracket is used.

It follows immediately from (\ref{i7}) that
\beq
\label{i10}
( N^{A}, N^{B} ) =\frac{1}{2}\left[ N^{A}\overleftarrow{\pa}_{C} ( N^{C}, N^{B} ) -
(A\leftrightarrow B )(-1)^{(\varepsilon_{A} + 1 )( \varepsilon_{B} + 1)} \right]. 
\eeq Here, in (\ref{i5}), (\ref{i7}), (\ref{ad}), (\ref{i10}), the usual
antibracket, generated by the operator $\Delta$, is used
\beq
\label{i8} ( f, g ) = (-1)^{\varepsilon(f)}[ [ \Delta, f ], g ]
\cdot 1=f \overleftarrow{\pa}_{A} E^{AB} \overrightarrow{\pa}_{B} g .      
\eeq
Thus, the coefficients $N^{A}$ of the vector field $N$ should satisfy Eqs.
(\ref{i6}) and (\ref{i7}).
Of course,  the simplest solution is obvious,
\beq
\label{i9}
N^{A} = Z^{A},   \quad  N = Z^{A}\pa_{A}.  
\eeq
That was exactly the simplest ansatz  used in the article \cite{BB2} from the very
beginning. Here, in the present article, we do not restrict ourselves with any {\it a priori}
choice of a special solution to Eqb. (\ref{i6}) and (\ref{i7}). Only these equations
themselves will be used in our further reasoning, nothing else.
It can be shown that the general solution to Eq. (\ref{i7}) is
\beq
\label{i11}
N^{A} = Z^{A} + 2( F, Z^{A} ), \quad   \varepsilon( F ) = 1       
\eeq
with $F$ being arbitrary Fermion.
In that case, Eqs. (\ref{i6}) and (\ref{i7}) are satisfied as follows:
\beq
\label{i12}
E^{AB} =\frac{1}{2}\left[ 2 E^{AB} + 2(F, E^{AB} ) \right] = E^{AB}, 
\eeq
\beq
\label{i13}
\Delta N^{A} = 2(\Delta F, Z^{A} ) = 0,   
\eeq
\beq
\label{i14}
\Delta F = {\rm const}( Z ).       
\eeq
Thus, we find
\beq
\label{i15}
N = Z^{A}\pa_{A} + 2\;{\rm ad}( F ).   
\eeq

One can rewrite Eq. (\ref{i7}) in its natural form
\beq
\label{i16}
\delta^{A}_{D} - \frac{1}{2} N^{A} \overleftarrow{\pa}_{D} =
\frac{1}{2} E^{AB} (\overrightarrow{\pa}_{B} N^{C } ) E_{CD},        
\eeq
where $E_{AB}$ is the inverse to $E^{AB}$,
\beq
\label{i17}
E^{AB} E_{BC} = \delta^{A}_{C}.    
\eeq
Now, the super trace of (\ref{i16}) yields
\beq
\label{i18}
\sTr I = \delta^{A}_{A} (-1)^{\varepsilon_{A}} = 0,   
\eeq
that is fulfilled identically due to equal  number of Bosons and Fermions
among the variables $Z^{A}$.
The supertrace imposes no restrictions for the divergence of vector field $N$,
\beq
\label{i19}
\div N = \pa_{A} N^{A} (-1)^{\varepsilon_{A}},       
\eeq
Note that $E^{AB}$ and $E_{AB}$ on the right-hand side of
(\ref{i16}) do enter in the form of a similarity transformation (
that is just the meaning  of the naturalness of (\ref{i16}) ) and,
therefore, they cancel each other when taking the supertrace
or superdeterminant. Notice also, that (\ref{i16}) is a "bridge"
between the initial equation (\ref{i7}) and its "dual" form, \beq
\label{i20} E_{AB} = \frac{1}{2} \left[ E_{AC} ( N^{C}
\overleftarrow{\pa}_{B} ) -
( A \leftrightarrow B ) (-1)^{ \varepsilon_{A} \varepsilon_{B} } \right].   
\eeq
In turn, by substituting
\beq
\label{i20.7}
N^{A} = - 2 E^{AB} V_{B}
 = 2 V_{B}  E^{BA},  
\eeq
\beq
\varepsilon( V_{A} ) = \varepsilon_{A} + 1,   
\eeq into (\ref{i20}), the latter takes the form
\beq
\label{i20.8}
E_{AB}  = \pa_{A} V_{B} -
\pa_{B} V_{A} (-1)^{\varepsilon_{A}\varepsilon_{B}}, 
\eeq
which tells us that $V_{A}$ is just the antisymplectic potential,
generating a constant invertible
metric in its covariant components $E_{AB}$. Thereby, one realizes
that the arbitrariness in $N^{A}$
is generated by the natural geometric arbitrariness in the choice of
the antisymplectic potential.
It can be shown that the general solution to Eq. (\ref{i20.8}) is
\beq
\label{i.V}
V_{B}  = \frac{1}{2} Z^{A} E_{AB} + \pa_{B} F, 
\eeq
with $F$ being arbitrary Fermion, so that (\ref{i.V})
is consistent with (\ref{i20.8}), (\ref{i20.7}).
Of course, it follows from (\ref{i6}), (\ref{i20.7}), that  the antisymplectic potential
$V_{A}$ should satisfy the condition
\beq
\label{i20.9}
\Delta V_{A} = 0,    
\eeq
which is consistent with (\ref{i14}).
The relation (\ref{i20.8}) is invariant under the shift,
\beq
\label{i2.31}
V_{A} = V'_{A} + \pa_{A} F' ,\quad  \varepsilon( F' ) = 1. 
\eeq
On the other hand, we have,
\beq
\label{i2.32}
\Delta V_{A}=\Delta V'_{A}+\pa_{A} \Delta F'(-1)^{\varepsilon_A} = 0,    
\eeq
\beq
\label{i2.33}
N^{A} = N^{'A}  - 2 E^{AB} \pa_{B} F',  
\eeq
\beq
\label{i2.34}
\Delta N^{A} = \Delta N^{'A}  +
2 E^{AB} \pa_{B} \Delta F' (-1)^{\varepsilon_{A}} = 0,  
\eeq
\beq
\label{i2.35}
N = N' + 2 {\rm ad}( F' ) ,  
\eeq
\beq
\label{i2.36}
\div N = \div N' - 4 \Delta F' .    
\eeq
So, if one chooses in (\ref{i2.31}) - (\ref{i2.35}) for the $F'$ to satisfy the relation
\beq
\label{i2.37}
\div N' = \div N + 4 \Delta F' = 0,   
\eeq
then the new value (\ref{i2.37}) of the $\div N'$ is zero. Thereby, the new
operator
\beq
\label{i2.38}
N' = N - 2 {\rm ad}(F') = - N'^{T},   
\eeq is an antisymmetric one. The transposed operation is defined
via
\beq
\int [dZ] (A^TF)G=(-1)^{\varepsilon(A)\varepsilon(F)}\int [dZ]
F(AG). \eeq So far, the condition (\ref{i2.37}) seems to be the only
restriction on $F'$. However, let us consider the commutator \beq
\label{i2.39}
[ \Delta, N' ]=[ \Delta,N - 2 {\rm ad}( F' ) ] = 2 \Delta -2 {\rm ad}( \Delta F' ). 
\eeq
So, if we would like for the new operator $N'$ to maintain the relation
\beq
\label{i2.40}
[ \Delta, N' ] = 2 \Delta,    
\eeq
then there should be
\beq
\label{i2.41}
\Delta F' = {\rm const} (Z).    
\eeq
Due to the latter, it follows from (\ref{i2.34})
\beq
\label{i2.42}
\Delta N^{A} = \Delta N'^{A} = 0.    
\eeq
In turn, due to (\ref{i2.41}), it follows from (\ref{i2.37}) that
\beq
\label{i2.43}
\div N = {\rm const} (Z).    
\eeq
Thus, we see that the deviation from zero allowed for $\div N$ is not so
arbitrary. That is because  the condition (\ref{i2.42}) is rather restrictive.
We see that the new antisymmetric $N'$ does maintain all the basic
conditions (\ref{i2.40}) and (\ref{i2.42}), provided the condition (\ref{i2.43}) holds.
In what follows below, we do mean that our $N'$-operator is chosen just in its
antisymmetrical form (\ref{i2.38}), "from the very beginning".  For brevity,
in all further formulae we omit the prime of $N'$.

There exists a crucially important consequence of (\ref{i7}),
that the operator $( N - 2 )$ does differentiate the antibracket,
\beq
\label{i26}
( N - 2 ) ( f, g ) = ( ( N - 2 ) f, g ) + ( f, ( N - 2 ) g ).    
\eeq
That goes as follows,
\beq
\nonumber
&&N ( f, g ) = ( N f, g ) + ( f, N g) - f \overleftarrow{\pa}_{A} \left[ (
N^{A}\overleftarrow{\pa}_{C} ) E^{CB} -
( A \leftrightarrow B ) (-1)^{ ( \varepsilon_{A} + 1 ) ( \varepsilon_{B} + 1
) } \right]\overrightarrow{\pa} _{B} g  =\\
\label{i27}
&&=( N f, g ) + ( f, N g ) - 2 ( f, g ),    
\eeq
which is equivalent to (\ref{i26}).

\section{Extended Antibracket}

Now, let us consider the antibracket generated by the extended operator
$\Delta_{\tau}$ \cite{BB2},
\beq
\label{i28}
( F, G )_{\tau} = (-1)^{\varepsilon(F)}[ [ \Delta_{\tau}, F], G ]\cdot 1=t^{2} ( F, G ) +
( N_{\tau} F ) \pa_{\theta} G - F \overleftarrow{\pa}_{\theta} N_{\tau} G,      
\eeq
where $( F, G )$ is the ( usual ) antibracket (\ref{i8}) in the original $Z^{A}$ -sector,
although functions $F, G$ themselves,
standing for $f, g$, respectively, do depend on $t, \theta $ as well. $N_{\tau}$ is
defined in (\ref{i3}).  Due to (\ref{i26}), the
operator $( N_{\tau} - 2 )$ does differentiate the usual antibracket $( F, G )$,
as well. In fact, we will use the relation equivalent to that,
\beq
\label{i30}
N_{\tau} ( F, G ) = ( N_{\tau} F, G ) + ( F, N_{\tau} G ) - 2 ( F, G). 
\eeq
One can state that the extended antibracket (\ref{i28}) does satisfy the strong
Jacobi identity, provided  the usual
antibracket has that property. In particular, one assumes that the strong
Jacobi identity holds for any Boson $B$,
\beq
\label{i31}
( ( B, B ) , B ) = 0, \quad     \varepsilon( B ) = 0. 
\eeq In its general form, the Jacobi identity can be reproduced from
(\ref{i31}) via the differential polarization procedure. To do this,
one has to choose a specific form for $B$
\cite{B},
\beq
\label{i32}
B = \sum_{i = 1}^{3} m_{i} n_{i},  \quad  n_{1} = F,\;  n_{2} = G, \; n_{3} =H. 
\eeq
Then, the operator
\beq
\label{i33}
\pa_{1} \pa_{2} \pa_{3} (-1)^{ ( \varepsilon_{1} + 1 ) ( \varepsilon_{3}
+ 1 ) + \varepsilon_{2} },    
\eeq
should be applied to (\ref{i31}), where $\pa_{i}$ are partial $m_{i}$ -derivatives,
$\varepsilon_{i}$ are Grassmann parities,
\beq
\label{i34}
\varepsilon_{i} = \varepsilon( n_{i} ) = \varepsilon( m_{i} ).   
\eeq

It is our task now, to prove that the extended antibracket (\ref{i28}) satisfies
the compact form of the strong Jacobi identities,
\beq
\label{i35}
( ( B, B )_{\tau}, B )_{\tau} = 0,   \quad     \varepsilon( B ) = 0,  
\eeq
provided  similar compact form (\ref{i31}) holds. We have
\beq
\label{i36}
( B, B )_{\tau} = t^{2} ( B, B ) + 2 ( N_{\tau} B ) ( \pa_{\theta} B). 
\eeq
By substituting that in (\ref{i35}), one gets
\beq
\nonumber
&&(( B, B)_{\tau}, B )_{\tau} = t^{2}( t^{2}( B, B )+2(N_{\tau}B)(\pa_{\theta}B),B)+
( N_{\tau} ( t^{2} ( B, B ) +\\
\label{i37}
&&+
2( N_{\tau}B)(\pa_{\theta} B ) ))( \pa_{\theta} B ) -
( \pa_{\theta} ( t^{2} ( B, B ) + 2 ( N_{\tau} B ) ( \pa_{\theta} B )
) ) ( N_{\tau} B ),    
\eeq
where the relations (\ref{i30}), (\ref{i31}) will be used, together with
\beq
\label{i38}
( \pa_{\theta} B )^{2} = 0     
\eeq
Thus, the right-hand side of (\ref{i37}) takes the form
\beq
\nonumber
&&( 2 t^{2} ( B, B ) + 2 t^{2} ( N_{\tau} B, B ) - 2 t^{2} ( B, B ) )
( \pa_{\theta} B ) +
2 t^{2} ( N_{\tau} B ) ( \pa_{\theta} B, B )-\\
\nonumber
 &&- 2 t^{2} ( N_{\tau}
B, B ) ( \pa_{\theta} B ) -
2 t^{2} ( \pa_{\theta} B, B ) ( N_{\tau} B ) + \\
\label{i39}
&&2 ( N_{\tau} B ) (
N_{\tau} \pa_{\theta} B ) ( \pa_{\theta} B ) -
2 ( \pa_{\theta} N_{\tau} B ) ( \pa_{\theta} B ) ( N_{\tau} B ) = 0.   
\eeq
Here, the first and third, the second and fifth, the fourth and sixth, the seventh and
eight terms,  compensate each other in
every pair mentioned.  Finally, the strong Jacobi identity (\ref{i35}) for the
extended antibracket (\ref{i28}) is proven.

\section{ Nontrivial Deformation in the Sector of Original Variables}

In turn, let us study the role of the operator $N$ in construction of a
nontrivially deformed antibracket in the original
$Z^{A}$-sector. So, let $\kappa$ be a deformation parameter. Consider the
operator
\beq
\label{i40}
K = K(N) = \kappa ( N - 2 ),\quad       K(N + 2) = \kappa N. 
\eeq
We have
\beq
\label{i41}
K(N + 2) - K(N) = 2 \kappa,     
\eeq
\beq
\label{i42}
K(N) ( f g )=( K(N + 2) f ) g + f( K(N) g)=( K(N) f ) g+f( K(N + 2)g ),      
\eeq
\beq
\label{i43}
K ( f, g ) = ( K f, g ) + ( f, K g ).     
\eeq
By using the well-known Witten formula for the usual antibracket,
\beq
\label{i44}
( f, g) = \Delta( f g ) (-1)^{\varepsilon(f)} - [ f (\Delta g) + (\Delta f )
g (-1)^{\varepsilon(f)} ],     
\eeq
we define the deformed antibracket by the relation \cite{BB2},
\beq
\label{i45}
( f, g )_{*} = \Delta ( f g ) (-1)^{\varepsilon(f)} -
( 1 - K ) [ f (\Delta_{*} g) + ( \Delta_{*} f ) g (-1)^{\varepsilon(f)} ] , 
\eeq
where
\beq
\label{i46}
\Delta_{*} = \Delta (1 - K )^{-1} = ( 1 - K( N + 2 ) )^{-1} \Delta. 
\eeq
By using (\ref{i42}) and (\ref{i44}), the relation (\ref{i45}) can be rewritten in the form
\beq
\label{i47}
( f, g )_{*} = ( f, g ) + ( K f ) ( \Delta_{*} g ) + ( \Delta_{*} f ) ( K g
) (-1)^{\varepsilon(f)}.      
\eeq
Usually, the deformation of the antibracket is defined by that formula.
It follows from (\ref{i45})
\beq
\label{i48}
\Delta_{*} ( f, g )_{*} = - \Delta \left[ f ( \Delta_{*} g) + ( \Delta_{*} f) g
(-1)^{\varepsilon(f)} \right].    
\eeq
\beq
\label{i49}
( ( \Delta_{*} f ), g )_{*} = \Delta ( ( \Delta_{*} f ) g)
(-1)^{\varepsilon(f) + 1} -
( 1 - K ) [ ( \Delta_{*} f ) ( \Delta_{*} g ) ],     
\eeq
\beq
( f, ( \Delta_{*} g ) )_{*} = \Delta ( f ( \Delta_{*} g ) )
(-1)^{\varepsilon( f )} -
( 1 - K ) [ ( \Delta_{*} f ) ( \Delta_{*} g ) ] (-1)^{\varepsilon( f )} , 
\eeq
\beq
\label{i50}
( ( \Delta_{*} f ), g )_{*} - ( f, ( \Delta_{*} g) )_{*}
(-1)^{\varepsilon(f)}  =
\Delta \left[ - ( \Delta_{*} f ) g (-1)^{\varepsilon(f)} - f ( \Delta_{*} g )
\right] = \Delta_{*} ( f, g )_{*}.     
\eeq
The latter means that the deformed operator (\ref{i46}) does differentiate the
deformed antibracket (\ref{i45}), or (\ref{i47}) \cite{BB2}.

Finally, one can state that the nontrivially deformed antibracket (\ref{i45}), or
(\ref{i47}), does satisfy the strong
Jacobi identity. Indeed, we have
\beq
\nonumber
&&( ( B, B )_{*}, B )_{*} = ( ( B, B ) + 2 ( K B ) ( \Delta_{*} B ), B ) +
( K ( ( B, B ) + 2 ( K B ) ( \Delta_{*} B ) ) ) ( \Delta_{*} B ) -\\
\nonumber
&&-
2 ( ( \Delta_{*} B ), B )_{*}  ( K B ) = 2 ( K B ) ( ( \Delta_{*} B ) , B
) - 2 ( ( K B ), B ) ( \Delta_{*} B ) +
2 ( ( K B ), B ) ( \Delta_{*} B )+\\
\label{i51}
&& + 2 ( K ( ( K B ) ( \Delta_{*} B ) ) ) (
\Delta_{*} B )  -
2 ( ( \Delta_{*} B ), B ) ( K B ) - 2 ( K \Delta_{*} B ) ( \Delta_{*} B )
( K B ) = 0.            
\eeq Here, on the right-hand side of the last equality, the first
and  fifth, the second and third, the fourth and sixth terms do
compensate each other in every pair mentioned.  Thus, the strong
Jacobi identity for the nontrivially deformed antibracket
(\ref{i45}), or (\ref{i47}), in the original $Z^{A}$ - sector is
proven.

It follows from (\ref{i48}), (\ref{i51}) that the operator
\beq
\label{i52a}
\sigma_{ * }( w )  =:  {\rm ad}_{ * }( w )  +
\frac{ \hbar }{ i } \Delta_{ *},    
\eeq
squared does satisfy the  equation
\beq
\label{i53a}
( \sigma_{ * }(w) )^{2}  =
{\rm ad}_{ * }\Big( \frac{1}{2} ( w, w )  +
\frac{ \hbar }{ i } \Delta_{ * } w \Big).       
\eeq
Thus, if the expression in the parentheses of the ${\rm ad}_{* }$ on the right-hand
side in (\ref{i53a}) is zero, then the operator (\ref{i52a}) is nilpotent.
The respective relations
in the non-deformed formalism are well known.

\section*{5\;\; Trivial $\tau$-Extended Deformation
\footnote{This  section represents the main  results and formulae of
Ref. \cite{BB2}, related in general  to  trivial deformations
in $\tau$ -extended phase space.}}
\setcounter{section}{5}
\renewcommand{\theequation}{\thesection.\arabic{equation}}
\setcounter{equation}{0}

Now, we have to consider a trivial $\tau$ extended deformation in the extended
phase space including
the variables $t$ and $\theta$. Let us introduce the operator
\beq
\label{i52}
 K_{\tau} = \kappa N_{\tau},     \quad    [ K_{\tau}, \Delta_{\tau} ]= 0, 
\eeq
and define trivially deformed  extended operator
\beq
\label{i53}
\Delta_{\tau *} = \Delta_{\tau} ( 1 - K_{\tau} )^{-1},
\quad \Delta^2_{\tau *} = 0.      
\eeq
Now, introduce the operator
\beq
\label{i54}
T = 1 + \kappa \theta \Delta_{\tau *} ,              
\eeq
and its inverse
\beq
\label{i55}
T^{-1} = 1 - \kappa \theta \Delta_{\tau}.                 
\eeq
Then, one finds that
\beq
\label{i56}
\Delta_{\tau *}  = T^{-1} \Delta_{\tau} T.      
\eeq
Also, it follows that the operator T does satisfy the equations (see also App. D)
\beq
\label{5.6}
[ \Delta_{\tau}, T ] = \Delta_{\tau} T K_{\tau}, \quad [ T, K_{\tau} ] =0. 
\eeq
Together with (\ref{i1}) and (\ref{i52}), these equations constitute what we
call "T-algebra".

Next, define a trivially deformed extended antibracket,
\beq
\nonumber
&&( F, G )_{\tau *} = T^{-1} ( ( T F ), ( T G ) )_{\tau} = \\
\nonumber
&&=( F, G
)_{\tau} +
( K_{\tau} F ) ( \Delta_{\tau *} G ) + ( \Delta_{\tau *} F )
(K_{\tau} G ) (-1)^{\varepsilon(F)} =\\
\label{i57}
&&=\Delta_{\tau} ( F G ) (-1)^{\varepsilon(F)} - ( 1 - K_{\tau} )\left [
F ( \Delta_{\tau *} G ) +
( \Delta_{\tau *} F ) G (-1)^{\varepsilon(F)} \right]. 
\eeq
The latter formula allows for a natural rewriting in terms of the $*$-
modified double-commutator formula generalizing (\ref{i8}) and (\ref{i28}),
\beq
\label{i57.3}
( F, G )_{\tau *}  =  (-1)^{\varepsilon(F)} [ [ \Delta_{\tau *},
( T F)_{*} ], ( T G )_{*} ]\cdot 1 ,    
\eeq
where we have used (\ref{i28}) and (\ref{i56}), and
\beq
\label{i57.4}
( T F )_{*}  =  T^{-1} ( T F ) T,  \quad   ( T G )_{*}  =  T^{-1} ( T G ) T.   
\eeq
Here in (\ref{i57.4}), on the right-hand sides, the operator $T$  in the middle
factors applies only to the function
standing to the right within the respective round bracket.

Notice that within the class of functions,
\beq
\label{i57.1}
 F = t^{-2} f, \quad  G = t^{-2} g,    
\eeq
the trivially deformed $\tau$ -extended operator (\ref{i56}) and antibracket (\ref{i57})
reduces, respectively,  to the
non-trivially deformed operator (\ref{i46}) and antibracket (\ref{i45})
in the original sector,
\beq
\label{i57.2}
\Delta_{\tau *}F=\Delta_{*}f,\quad ( F, G )_{\tau *} = t^{-2} ( f, g )_{*} .   
\eeq

By construction, the trivially deformed extended antibracket (\ref{i57}) does
satisfy the strong Jacobi identity.
In turn, define a trivial associative and commutative  star-product
\beq
\label{i58}
( F * G ) = T^{-1} ( ( T F ) ( T G ) ) = F G - \kappa \theta ( F, G
)_{ \tau *} (-1)^{\varepsilon(F)}.     
\eeq
It is worthy to mention here that the operators (\ref{i57.4}) apply to a function
as to yield the left adjoint of the symbol multiplication (\ref{i58}),
\beq
\label{i58.1}
( (T F)_{*} G ) =  F * G,\quad (T F)_{*} = F -
\kappa\!\; \theta\!\; {\rm ad}_{\tau_{*}} ( F ) (-1)^{\varepsilon( F )}.  
\eeq

 Due to (\ref{i57.2}), within the class of functions (\ref{i57.1}),
 the star-product (\ref{i58}) reduces as follows
\beq
\label{5.13}
F * G = ( t^{-2} f ) ( t^{-2} g ) - \kappa \theta t^{-2} ( f, g )_{*}
(-1)^{\varepsilon(f)}.    
\eeq
Then, with respect to the star-product (\ref{i58}), we have the
trivially  deformed extended Witten formula \footnote{The same as in
the undeformed case, the deformed extended Witten formula
(\ref{i59}) follows directly from the double-commutator formula
(\ref{i57.3}) with (\ref{i57.4}), (\ref{i58}), (\ref{i58.1}) taken
into account. } \beq \label{i59} ( F, G )_{\tau *} = \Delta_{\tau *}
( F * G ) (-1)^{\varepsilon(F)} - F
* ( \Delta_{\tau *} G ) -
( \Delta_{\tau *} F ) * G (-1)^{\varepsilon(F)},     
\eeq
the Leibnitz rule,
\beq
\label{i60}
( ( F * G ), H )_{\tau *} =  F * ( G, H )_{\tau *} +
G * ( F, H )_{\tau *} (-1)^{\varepsilon(F) \varepsilon(G)}, 
\eeq
the Getzler identity \cite{G} providing for the absence of higher antibrackets  in
the BV -algebra
\beq
\nonumber
&&\Delta_{\tau *} ( F * G * H ) - \Delta_{\tau *} ( F * G ) * H -
F * \Delta_{\tau *} ( G * H ) (-1)^{\varepsilon(F)} -\\
\nonumber
&&- \Delta_{\tau *} ( F * H ) * G (-1)^{\varepsilon(G)
\varepsilon(H)}+( \Delta_{\tau *} F ) * G * H-\\
\label{i61}
&&- F * ( \Delta_{\tau *} G )* H
(-1)^{\varepsilon_(F)} +
F * G * ( \Delta_{\tau *} H ) (-1)^{\varepsilon(F) +
\varepsilon(G)} = 0.     
\eeq
The star exponential  is defined as (see also App. C)
\beq
\nonumber
&& \exp_{*}\{ B \}  =
1 + B + \frac{1}{2} B * B + \frac{1}{3!} B * B * B + ... =\\
\label{i62}
&&=T^{-1} \exp\{( T B )\} =
\exp\left\{ B -\frac{1}{2} \kappa \theta ( B, B )_{\tau *} \right\}.    
\eeq
The latter satisfies
\beq
\label{i63.1}
&&\exp_{*}\{ - B \}  *  \exp_{*}\{ B \}  = 1, \\
\label{i63.2}
&&\exp_{*}\{ - B \}  * (
\Delta_{\tau *} \exp_{*}\{ B \} ) =
( \Delta_{\tau *} B ) + \frac{1}{2} ( B, B )_{\tau *}, \\
\label{i63.3}
&&\exp_{*}\{ B + B'\} =\exp_{*}\{ B \}  * \exp_{*}\{ B'\}, \\
\label{i63.4}
&&\delta  \exp_{*}\{ B \}  = \exp_{*}\{ B \} * \delta B,\quad
\varepsilon(B) = \varepsilon(B') = 0.              
\eeq

As for the reduced $B$-form,
\beq
\label{i63.5}
B  =:  t^{-2} \frac{ i }{ \hbar } w,    
\eeq
due to the above relations (\ref{i57.2}), the one (\ref{i63.2}) yields exactly
the expression in the parentheses of the ${\rm ad}_{*}$ on the right-hand side
in (\ref{i53a}),
\beq
\label{i63.6}
\Big( \exp_{*}\Big\{ - t^{-2} \frac{ i }{ \hbar } w \Big\} * \Delta_{ \tau * }
\exp_{*}\Big\{  t^{-2} \frac{ i }{ \hbar } w \Big\} \Big)_{ t = 1, \theta = 0 } =
\Big( \frac{i}{\hbar} \Big)^{2} \Big( \frac{1}{2} (w, w)_{*} + \frac{\hbar}{i}
\Delta_{*} w \Big).
\eeq

The trivially deformed extended  quantum master equation has the form
\beq
\label{i64}
\Delta_{\tau *} \exp_{*}\left\{ \frac{i}{\hbar}\; W \right\}  = 0, 
\eeq
or, equivalently,
\beq
\label{i65}
\frac{1}{2} ( W, W )_{\tau *} = i \hbar \Delta_{\tau *} W.     
\eeq
One has to seek for a solution to that equation in the form
\beq
\label{i5.21}
W = \sum_{k = - 2}^{\infty} W_{ (k|0) } t^{k} + \theta \sum_{k = 1}^{\infty}
W_{ (k|1) } t^{k},    
\eeq
where the component $W_{ (-2|0) } = S$  is  identified with the classical
nontrivially  deformed proper action, (see also App. B)
\beq
\label{i5.22}
( S, S )_{*} = 0.     
\eeq
The detailed form of the equations for coefficients in (\ref{i5.21}), together with
the corresponding formal techniques, can be found in Ref. \cite{BB2}.

The trivially deformed  path integral with  a measure $d\mu$  in the extended
phase space is defined as
\beq
\label{i66}
\mathcal{Z} = \int d\mu \exp_{* ( \kappa)}\left\{\frac{i}{ \hbar}\; W \right\}
 \exp_{* ( -\kappa)}\left\{\frac{i}{\hbar }\; X \right\}   =
\int d\mu \exp\left\{\frac{i}{\hbar }\; A \right\}, 
\eeq
where
\beq
\label{5.27}
A = T_{ (\kappa)} W + T_{( - \kappa)}X,         
\eeq
\beq
\Delta_{\tau * ( \kappa ) }\left(\exp_{*(\kappa )}\left\{\frac{ i}{ \hbar}\; W \right\}
\right) = 0,\quad
\Delta_{\tau * (- \kappa )}\left(\exp_{*( - \kappa )} \left\{\frac{i}{\hbar}\; X \right\}
\right)=  0. 
\eeq
Here, $X$ satisfies the same equation as $W$ does, but with the formal replacement
$\kappa \rightarrow - \kappa$.
This replacement just provides for right transposition properties when
integrating by part.

We proceed in (\ref{i66}) with the following integration measure
\beq
\label{i5.25}
d \mu = t^{-1} d t d\theta d \lambda_{\theta} [ d Z ] [ d \lambda ].    
\eeq
The transposed operator $A^{T}$ of the operator $A$ is defined via
\beq
\label{i5.26}
\int d\mu ( A^{T} F ) G  =   (-1)^{\varepsilon(A) \varepsilon(F)}  \int d
\mu  F  ( A G).       
\eeq
Our main transposed operators are
\beq
\label{i5.27}
\Delta^{T} = \Delta,  \quad N^{T} = - N, \quad
\Delta^{T}_{\tau} = \Delta_{\tau},\quad N^{T}_{\tau} = - N_{\tau},\quad
\Delta^{T}_{\tau *( \kappa )} =  \Delta_{\tau * (- \kappa )}.   
\eeq
Let us make in (\ref{i66}) the variation of the form
\beq
\label{i5.28}
\delta \exp_{*(-\kappa)}\left\{\frac{ i}{\hbar }\; X \right\} =
\Delta_{\tau *(-\kappa)} \left( \exp_{*(-\kappa)}\left\{\frac{i}{\hbar
}\; X \right\} *_{(-\kappa)} \delta \Psi \right),    
\eeq
with arbitrary infinitesimal Fermion $\delta \Psi$. The (\ref{i5.28}) is consistent
with (\ref{i63.4}) due to the quantum master equation for the  $X$.
Then, we deduce that the path
integral is independent of the gauge-fixing action $X$,
\beq
\nonumber
&&\delta_{X} \mathcal{Z} = \int d \mu \exp_{*( \kappa )}\left\{\frac{i}{\hbar }\;W \right\}
\delta \exp_{* ( -\kappa ) }\left\{\frac{i}{\hbar }\; X \right\} =\\
\nonumber
&&=\int d \mu \exp_{* ( \kappa ) }\left\{\frac{i}{\hbar}\; W \right\}\Delta_{\tau *( -
\kappa )} \left( \exp_{ * ( - \kappa ) }\left\{\frac{i}{\hbar }\; X \right\}
*_{ ( - \kappa ) } \delta \Psi \right) =\\
\label{i5.29}
&&=\int d \mu \left( \Delta_{\tau * ( \kappa )}
\exp_{ * ( \kappa ) }\left\{\frac{i}{\hbar }\; W \right\} \right)
\left(\exp_{* ( - \kappa) }
\left\{\frac{i}{\hbar }\; X \right\} *_{ ( - \kappa ) } \delta \Psi \right)
= 0.      
\eeq

It follows from (\ref{i56}), (\ref{i62}) and (\ref{i64}) that
\beq
\Delta_{\tau}  \exp\left\{\frac{i}{\hbar }\; T_{ ( \kappa) } W  \right\} = 0.
\eeq
For similar reasons, it follows that
\beq
\Delta_{\tau}\exp\left\{\frac{i}{\hbar}\;T_{( - \kappa ) } X \right\} = 0.
\eeq
These equations tell us that in terms of the barred actions,
\beq
\bar{W} = T_{ (\kappa) } W,\quad \bar{X} = T_{ ( -\kappa ) } X,
\eeq
the path integral (\ref{i66}) is just the standard $W - X$ version of the
field-antifield formalism. From the latter point of view, it is
well-known that the path integral (\ref{i66}) is stable  under the
gauge variation
\beq
\delta{ \bar{X} } = \sigma_{\tau}(\bar{X}) \bar{\Psi },
\eeq
where
\beq
\sigma_{\tau}( \bar{X} ) =-i\hbar \;\Delta_{\tau}+{\rm ad}_{\tau}( \bar{X} ),
\eeq
so that
\beq
\delta X = T^{-1}_{( - \kappa ) } \delta \bar{X}.
\eeq
 If one identifies $\bar{ \Psi } =T_{ ( - \kappa ) } \Psi$ , then
\beq
\delta X = \sigma_{ \tau * ( -\kappa ) }( X ) \Psi ,
\eeq
where
\beq
\sigma_{ \tau * ( - \kappa ) }( X )
= -i\hbar\; \Delta_{\tau *( - \kappa )}+{\rm ad}_{\tau *( -\kappa)}( X ),
\eeq
which is exactly the variation of $X$ generated by (\ref{i5.28}).

\section{ Gauge-Fixing in the Classical Extended Nondeformed\\ Master
Equation}

Here we study, if the standard gauge-fixing procedure is capable to
eliminate the extra variable $t$, as applied
to the classical  $\tau$ -extended nondeformed master equation,
\beq
\label{i6.1}
( S, S )_{\tau} = t^{2} ( S, S ) + 2 ( N_{\tau} S ) ( \pa_{\theta} S ) = 0. 
\eeq
where we restrict ourselves to the simplest choice for $N_{\tau}$,
\beq
\label{i6.2}
N_{\tau} = Z^{A} \pa_{A} + t \pa_{t}.    
\eeq
We proceed with the following ansatz for S,
\beq
\label{i6.3}
S  = S ( Z, t, \theta ) = S ( t^{-1} Z, \theta  ),        
\eeq
which implies
\beq
\label{i6.4}
N_{\tau} S = 0,     
\eeq
so that Eq.  (\ref{i6.1}) takes the usual form of the classical master
equation,
\beq
\label{i6.5}
( S, S ) = 0.       
\eeq
Let $\mathcal{S}( \phi )$ be an original action of original fields $\phi^{i}$ ,
and $R^{i}_{\alpha}( \phi )$ do satisfy the Noether identities,
\beq
\label{i6.6}
\mathcal{S} \overleftarrow{\pa}_{i} R^{i}_{\alpha} = 0,
\quad \pa_i=\frac{\pa}{\pa \phi^i}.     
\eeq
For the sake of simplicity, let the generators $R^{i}_{\alpha}$ be
linearly independent, so that the theory is irreducible.  Let us expand
the ansatz (\ref{i6.3}) in powers of antifields,
\beq
\nonumber
&&S = \mathcal{S}(t^{-1}\phi) + t^{-1}\phi^*_{i} R^{i}_{\alpha}(t^{-1}\phi)
t^{-1} C^{\alpha} + \theta t^{-1} C^{t} +
\frac{1}{2}t^{-1} C^*_{\gamma} U^{\gamma}_{\alpha \beta}(t^{-1}\phi) t^{-1}
C^{\beta} t^{-1} C^{\alpha} (-1)^{\varepsilon_{\alpha}} +\\
\label{i6.7}
&&+
t^{-1} {\bar C}^{*\alpha} t^{-1} B_{\alpha} + t^{-1} {\bar C}^{*t} t^{-1}
B_{t} + ... ,       
\eeq
where the terms presented explicitly are enough for a rank one theory, while
ellipses mean terms nonlinear in antifields.

Now, let us split $Z^{A}$ into fields and antifields,
\beq
\label{i6.8}
Z^{A} = \{\Phi^{a}, \Phi^*_{a}\}.  
\eeq
Then the gauge-fixing Fermion allowed has the form
\beq
\label{i6.9}
\Psi= \Psi( \Phi, t ) = \Psi( t^{-1} \Phi, \ln t ),       
\eeq
so that the antifields $\Phi^*_{a}$ should be eliminated in (\ref{i6.7}) by the
conditions
\beq
\label{i6.10}
\Phi^*_{a} = t ^{2} \Psi \overleftarrow{\pa}_{a}, \quad \pa_a=\frac{\pa}{\pa \Phi^a},     
\eeq
\beq
\label{i6.11}
\theta = N_{\tau} \Psi.        
\eeq
These conditions do correspond to the following ansatz for the gauge-fixing
master action X,
\beq
\label{i6.12.1}
X = \left( t^{-1} \Phi^*_{a} - t \Psi \overleftarrow{\pa}_{a} \right) \lambda^{a} + (
\theta - N_{\tau} \Psi ) \lambda^{\theta},      
\eeq
where $\lambda^{a}$ and $\lambda^{\theta}$ is the corresponding Lagrange
multiplier.
In fact, it is enough for our purposes to use the ansatz
\beq
\label{i6.12}
\Psi = t^{-1} \bar{C}_{\alpha} \chi^{\alpha}( t^{-1} \phi ) + t^{-1}
\bar{C}_{t} \chi^{t}( \ln t ),    
\eeq
with the following identification of fields
\beq
\label{i6.13}
\Phi^{a} = \{\phi^{i}, B_{\alpha}, C^{\alpha}, \bar{C}_{\alpha}, B_{t}, C^{t},
\bar{C}_{t}\}.    
\eeq

Thus, we arrive at the following complete gauge-fixed action
\beq
\nonumber
&&S_{gauge-fixed} = \mathcal{S}( t^{-1}\phi ) +
 t^{-1} \bar{C}_{\alpha} \chi( t^{-1} \phi ) \overleftarrow{\pa}_{i}
R^{i}_{\alpha}( t^{-1} \phi ) C^{\alpha} +
t^{-2} \bar{C}_{t}( N_{\tau} \chi^{t} ) C^{t}+ \\
\label{i6.14}
&&+ \chi^{\alpha}( t^{-1} \phi
) t^{-1} B_{\alpha}  + \chi^{t}( \ln t ) t^{-1} B_{t} + ... .   
\eeq
That action has, in its terms presented explicitly, the standard structure
of the Faddeev-Popov action, both
in the sector of the usual gauge $\chi^{\alpha}$ and of the extra gauge
$\chi^{t}$.  By choosing $\chi^{t} = \ln t$,
one removes the $t$-integration at the value $t = 1$. Thereby, it is shown that
the extra variable t is eliminated actually via the standard gauge-fixing procedure.

\section{ Gauge-Fixing in the Extended Trivially Deformed Classical/Quantum Master
Equation}

Let us consider the  extended trivially deformed classical master equation,
\beq
\label{i7.1}
( S, S )_{\tau{*}} = 0,     
\eeq
or in more detail,
\beq
\label{i7.2}
t^{2} (S, S ) + 2 ( N_{\tau} S )( \pa_{\theta} S ) +
2 \kappa ( N_{\tau} S ) ( ( t^{2} \Delta + N_{\tau} \pa_{\theta} )
( 1 - \kappa N_{\tau} )^{-1} S ) = 0.     
\eeq
The same as in  Sec. 6, we have chosen $N _{\tau}$ in the
simplest form (\ref{i6.2}).
In contrast to the previous section, we are not allowed to require for the
operator $N_{\tau}$
to annihilate the $S$, as the deformation by itself would be eliminated
immediately in this way.
However, if we do believe that a solution for $S$ does exist, we can try to
require for $N_{\tau}$ to
annihilate the gauge-fixing part of $S$, at least. When doing that, we should
provide for the form
of Eqs. (\ref{i7.1}), or (\ref{i7.2}), to be respected. Let us seek for $S$ in the
form
\beq
\label{i7.3}
S = S_{min} + t^{-2} ( \bar{C}^{*{\alpha}} B_{\alpha} + \bar{C}^{*{t}} B_{t} ),   
\eeq
where the minimal action $S_{min}$ depends on the minimal,  gauge-algebra
generating, set of variables, only,
\beq
\label{i7.4}
S_{min} = S_{min}( \phi, \phi^*; t, \theta; C, C^*; C^{t}, C^*_{t} ).   
\eeq
By construction, the second and third term in (\ref{i7.3}), those are just the
gauge-fixing parts of $S$ , are
certainly annihilated by the $N_{\tau}$, while the $S_{min}$ is not. In this
way, one can see that
the $S_{min}$ by itself does satisfy exactly Eqs. (\ref{i7.1}),
or (\ref{i7.2}). If
one chooses the gauge
Fermion $\Psi$ in the simplest form (\ref{i6.12}), then the antifields should be
eliminated by the conditions
(\ref{i6.10}), see also (\ref{i6.13}).  In turn, the second and third term
in (\ref{i7.3})
take exactly the form of fourth
and fifth term in (\ref{i6.14}), respectively. Thereby, it is shown that the extra
variable $t$  is eliminated
actually via the standard gauge-fixing procedure.

In the same way, one can consider the extended trivially deformed quantum
master equation, (\ref{i65}),
\beq
\label{7.5}
t^{2} ( W, W ) + 2 ( N_{\tau} W ) ( \pa_{\theta} W ) +
2( (\kappa N_{\tau} W)  - i \hbar) (( t^{2} \Delta + N_{\tau}
\pa_{\theta} ) ( 1 - \kappa N_{\tau} )^{-1} W ) =  0.     
\eeq
As Eq. (\ref{i7.2}) is a classical limit to the quantum equation (\ref{7.5}),
all the above reasoning, as well as the final statement remains the same.

Finally, let us notice the following. In Secs. 6 and 7,  we have used the
zero mode $t^{-1} Z^{A}$ of the
operator (\ref{i6.2}).  Here, we mention in short how to deal with the general
operator (\ref{i3}), where $N^{A}$
is defined by (\ref{i20.7}) - (\ref{i20.9}). Let $\bar{Z}^{A}$
be the zero mode of the
operator (\ref{i3}). Then we have
formally,
\beq
\label{i7.5}
\bar{Z}^{A} = \exp\{ - ( \ln t ) N \} Z^{A},  \quad    N = N^{A} \pa_{A}. 
\eeq
It follows from (\ref{i26}) that
\beq
\label{i7.6.1}
( \bar{Z}^{A}, \bar{Z}^{B} ) = t^{-2} E^{AB}, \quad
( \bar{Z}^{A}, Z^{C} ) E_{CD} ( Z^{D}, \bar{Z}^{B} )=( \bar{Z}^{A},\bar{Z}^{B} ).  
\eeq
In  turn,  it  follows  from  (\ref{i7.6.1})  that  the general  solution for  the
left  off-diagonal  block  has  the  form,
\beq
\label{78}
( \bar{Z}^{A}, Z^{B} )  =  t^{-1} S^{A}_{\;\;C}( t ) E^{CB}, \quad
\bar{Z}^{A}  =  t^{-1} S^{A}_{\;\;B}( t ) Z^{B}, 
\eeq
where $S^{A}_{\;\;B}(t)={\rm const}(Z)$ is a $t$-dependent antisymplectic matrix,
\beq
\label{7.9}
S^{A}_{\;\;C}( t ) E^{CD} S^{B}_{\;\;D}( t ) (-1)^{\varepsilon_{D} ( \varepsilon_{B}
+ 1 ) }  =  E^{AB} ,     
\eeq
such that
\beq
\label{7.10}
S^{A}_{\;\;B}( t = 1 ) = \delta^{A}_{\;\;B}.        
\eeq
One can get the general solution for the right off-diagonal block via the
supertransposition  in (\ref{78}).

Now, let us consider some explicit formulae concerning the modified
$N$-operator. First, let us choose the quadratic Fermion $F$ entering (\ref{i11})
that meets the condition (\ref{i14}),
\beq
\label{7.11}
2 F = Z^{A} F_{AB} Z^{B},  \quad   \varepsilon (F) = 1,     
\eeq
\beq
\label{7.12}
\varepsilon( F_{AB} )  = \varepsilon_{A}+\varepsilon_{B}+1,  \quad
F_{AB} = F_{BA} (-1)^{\varepsilon_{A} \varepsilon_{B}} = {\rm const}(Z).   
\eeq
We have
\beq
2\Delta F = E^{AB} F_{BA} (-1)^{ \varepsilon_{A} } = {\rm const}( Z ), 
\eeq
\beq
\label{7.14}
N^{A} = Z^{A} + 2( F, Z^{A} ) = Z^{B} ( \delta_{B}^{\;\;A} + 2F_{BC} E^{CA} ) =
(\delta^{A}_{\;\;B} - 2 E^{AC} F_{CB} ) Z^{B}.     
\eeq
On the other hand,  as the second in (\ref{78})  is the zero mode of $N_{\tau}$,
we have another expression for $N^{A}$,
\beq
\label{7.15}
N^{A} = - ( S^{-1} )^{A}_{\;\;C}\; t^{2} \pa_{t} ( t^{-1} S^{C}_{\;\;B} ) Z^{B}.
\eeq
It follows from  (\ref{7.14}) and (\ref{7.15}) that the Lie equation holds
\beq
\label{7.16}
t \pa_{t} S^{A}_{\;\;B} = 2 S^{A}_{\;\;C} E^{CD} F_{DB}, \quad
S^{A}_{\;\;B}( t = 1 ) =\delta^{A}_{\;\;B},      
\eeq
whose formal matrix solution is
\beq
\label{7.17}
S =  S( t ) = \exp\{ 2\ln( t ) E F \}.      
\eeq
By $t$-differentiating the formula (\ref{7.9}), and then using the equation (\ref{7.16}),
one confirms that  the matrix (\ref{7.17})
by  itself  does satisfies exactly  the antisymplicticity equation (\ref{7.9}).
Thus, we see that the exponential (\ref{7.17})
provides for the  $t$-parametrization of a family of antisymplectic matrices,
with $( EF )$ being a generator.

If one splits the full set  $Z^{A}$ into minimal sector $Z_{min}$ in (\ref{i7.4}),
except  for $\{ t, \theta \}$, and
the rest, $Z_{aux}$, $\{Z\} =\{ Z_{min}\}\oplus \{Z_{aux}\}$, then, by choosing in (\ref{i.V})
$F=F(Z_{min})$,  one has
\beq
\label{i7.6.2}
\bar{Z}_{aux} = t^{-1} Z_{aux}. 
\eeq
In terms of (\ref{i7.5}), the formula (\ref{i6.3}) and (\ref{i6.9}) takes the form,
\beq
\label{i7.6}
S = S( \bar{Z}, \theta ),    
\eeq
and
\beq
\label{i7.7}
\Psi = \Psi( \bar{\Phi}, \ln t ),    
\eeq respectively.
In turn, the formula (\ref{i7.3}) in terms of
(\ref{i7.5}) preserves its 
form
\beq
\label{i7.8}
 S = S_{min} +
(\overline{\bar{C}^{*\alpha}B_{\alpha}}   +
\overline{\bar{C}^{*t}B_t } )=
S_{min} + t^{-2} ( \bar{C}^{*{\alpha}} B_{\alpha} + \bar{C}^{*{t}} B_{t} ),     
\eeq
where $S_{min}$ is given by (\ref{i7.4}).
For the particular case $N^{A} = Z^{A}$, we reproduce from (\ref{i7.5})
$\bar{Z}^{A} = t ^{-1} Z^{A}$.

Due to (\ref{i7.6.2}) and (\ref{i7.8}),  one has
\beq
\label{i7.11}
( S_{min}, S_{aux} ) = 0,  \;\;  ( S_{aux}, S_{aux} ) = 0,  \;\; N_{\tau} S_{aux} =
0, \;\;  \pa_{\theta} S_{aux} = 0,\;\; S_{aux} = S - S_{min},    
\eeq
together with
\beq
\label{i7.12}
\Delta S_{aux} = 0.    
\eeq
The relations (\ref{i7.11}) and (\ref{i7.12}) allow one to preserve the form of the equation
(\ref{i7.2}) for the minimal action (\ref{i7.4}) in the general case of (\ref{i7.5}).

\section{ Generalized Darboux Coordinates \cite{BB2}}

The $\tau$ -extended trivially deformed classical/quantum master
equation takes its simplest form  in the so-called generalized
Darboux coordinates,
\beq
\label{i8.1}
\tau_{0}:  Z^{A}_{0} = t^{-1} Z^{A},\quad t_{0} = \ln t,\quad  t^*_{0} = \theta, 
\eeq
with the following integration measure,
\beq
d\mu=d t_{0} d t^*_{0} d\lambda_{t^*_{0}} [ d Z_{0} ] [ d \lambda ].
\eeq
We have already used these coordinates  partially when discussing
the gauge-fixing
procedure. In the case of the master equation, we have, with the use
of (\ref{i8.1}),
\beq
\label{i8.2}
( W, W )_{\tau_{0}{*}} = 2  i  \hbar \Delta_{\tau_{0}{*}} W,    
\eeq
where
\beq
\label{i8.3}
( F, G )_{\tau_{0}{*}} = ( F, G )_{\tau_{0}} + ( K_{\tau_{0}} F ) (
\Delta_{\tau_{0}{*}} G ) +
( \Delta_{\tau_{0}{*}} F ) ( K_{\tau_{0}} G ) (-1)^{
\varepsilon_{F} }, 
\eeq
\beq
\label{i8.4}
( F, G )_{\tau_{0}} = F [ \overleftarrow{\pa}_{A 0} E^{AB}
\overrightarrow{\pa}_{B 0} +
\overleftarrow{\pa}_{t_{0}} \overrightarrow{\pa}_{t^*_{0}} -
\overleftarrow{\pa}_{t^*_{0}} \overrightarrow{\pa}_{t_{0}} ] G,\quad   \pa_{A 0} =
\frac{\pa}{\pa Z^{A}_{0}},      
\eeq
\beq
\label{i8.5}
K_{\tau_{0}} = \kappa \pa_{t_{0}},   
\eeq
\beq
\label{i8.6}
\Delta_{\tau_{0}{*}} = \Delta_{\tau_{0}} ( 1 - K_{\tau_{0}} )^{-1}, 
\eeq
\beq
\label{i8.7}
\Delta_{\tau_{0}} = \frac{1}{ 2 } (-1)^{\varepsilon_{A}} \pa_{A 0} E^{AB}
\pa_{B 0} + \pa_{t_{0}} \pa_{t^*_{0}}.    
\eeq
We have the two main simplifications here. The first is the absence
of the $t_{0}$ -dependent factors in the square bracket in (\ref{i8.4}).  The second
is a very simple form of the  operator (\ref{i8.5}). The latter reduces to the $t_{0}$
-derivative.

The antifields are eliminated by the conditions
\beq
\label{i8.8}
\Phi^*_{a 0} = \Psi\overleftarrow{\pa}_{a 0},  \quad   t^*_{0} =
\Psi\overleftarrow{\pa}_{t_{0}},   
\eeq
where
\beq
\label{i8.9}
Z^{A}_{0} = \{\Phi^{a}_{0},  \Phi^*_{a 0}\},\quad
\pa_{a 0} = \frac{\pa}{\pa \Phi^{a}_{0}}.         
\eeq

Finally, let us consider in short what happens  if one ignores the
gauge-fixing mechanism as to eliminate the extra variable $t$.
In that case, one assumes the $W$ to be $\theta$ -independent,
\beq
\label{i8.10}
\pa_{\theta} W = \pa_{t^*_{0}} W = 0.     
\eeq
Under the assumption (\ref{i8.10}),  the path integral  (\ref{i66}) becomes as
represented in coordinates (\ref{i8.1}),
\beq
\label{i8.11}
\mathcal{Z}= \int d t_{0}\int d \Phi_{0}\exp \left\{\frac{i}{\hbar}A \right\}, 
\eeq
\beq
\label{i8.12}
A = W( \Phi, \Phi^*, t, \kappa, \hbar ) = W \left(\exp\{ t _{0} \} \Phi_{0},
\exp\{ t_{0} \} \left( \Psi( \Phi_{0} )\frac{\overleftarrow{\pa}}{\pa\Phi_{0}} \right),
\exp\{t_{0}\},  \kappa, \hbar \right),       
\eeq
\beq
\label{i8.13}
( W, W )_{0} + 2 \left( \kappa ( \pa_{t_{0}} W )  -  i \hbar  )
( \Delta_{0} ( 1 - \kappa \pa_{ t_{0} } )^{-1} W \right) = 0,     
\eeq
\beq
\label{i8.14}
( F, G )_{0} = F \overleftarrow{\pa}_{A 0} E^{AB}\overrightarrow{\pa}_{B 0} G, 
\eeq
\beq
\label{i8.15}
\Delta_{0} =\frac{1}{2 } (-1)^{\varepsilon_{A}} \pa_{A 0} E^{AB} \pa_{B 0}. 
\eeq

In Ref. \cite{BB2}, it was suggested that the extra variable $t_{0}$, remaining in the
path integral (\ref{i8.11}), plays the role of the Schwinger proper time
in the field-antifield formalism. It
seems rather plausible that the variable $t$ has a non-perturbative status. If
one rescales in (\ref{i8.12}):  $W \rightarrow \exp\{ - 2 t_{0} \} W$,
then in (\ref{i8.12}), (\ref{i8.13}) one should
substitute: $\hbar \rightarrow \exp\{ 2 t_{0} \} \hbar$,
$ \pa_{t_{0}}\rightarrow \pa_{t_0} - 2$.

\section*{Acknowledgments}
\noindent
 I. A. Batalin would like  to thank Klaus Bering of Masaryk
University for interesting discussions. The work of I. A. Batalin is
supported in part by the RFBR grants 14-01-00489 and 14-02-01171.
 The work of P. M. Lavrov is supported in part by the Presidential grant
 88.2014.2 for LRSS and by the RFBR grant 15-02-03594.
\\

\appendix
\section*{Appendix A. General Coordinates}
\setcounter{section}{1}
\renewcommand{\theequation}{\thesection.\arabic{equation}}
\setcounter{equation}{0}

Although the problem of a deformation of the antibracket on the general
antisymplectic manifold still has no mathematical
status established, it seems rather interesting to consider a formal
generalization of our basic equations to the case of general
antisymplectic coordinates. As a matter of some simple formal manipulations,
such a generalization appears quite natural
and "minimal", and looks very nice, as well.  Here, we present the
general-coordinate counterpart to the basic equation (\ref{i4}),
which provides, in turn, for the nilpotency of the extended  $\Delta$
-operator (\ref{i1}).

Let $Z^{A}$ be general coordinates on an antisymplectic manifold with
invertible antisymplectic metric $E^{AB}$, and measure density $\rho$.
Let these objects be
compatible in the sense that the odd Laplacian operator
\beq
\label{A1}
\Delta = \frac{1}{2} E^{B} \pa_{B} +\frac{1}{2} (-1)^{\varepsilon_{A}} E^{AB} \pa_{B}
\pa_{A},   
\eeq
\beq
\label{A2}
E^{B} = (-1)^{\varepsilon_{A}} \rho^{-1} ( \pa_{A} \rho E^{AB} ),      
\eeq
is nilpotent,
\beq
\label{A3}
\Delta^{2} = 0.  
\eeq

Let us consider our basic equation,
\beq
\label{A4}
[ \Delta, N ] = 2 \Delta,  \quad N = N^{A}\pa_{A} ,    
\eeq
which implies
\beq
\label{A5}
2 \Delta  N^{C} = ( N + 2 ) E^{C},     
\eeq
\beq
\label{A6}
( N^{A}, Z^{B} ) - ( A \leftrightarrow B ) (-1)^{ ( \varepsilon_{A} + 1 ) (
\varepsilon_{B} + 1 ) } = ( N + 2 ) E^{AB}.      
\eeq
These equations are general-coordinate counterparts to our Eqs. (\ref{i6}) and
(\ref{i7}).
It is a remarkable fact that Eqs. (\ref{i20.7}) and (\ref{i20.8}) remain valid in
the general coordinates  as well.
 Namely, it follows from (\ref{A6}) that its dual holds,
\beq
\label{A7}
( \pa_{A} N^{C} ) E_{CB} - ( A \leftrightarrow B ) (-1)^{\varepsilon_{A}
\varepsilon_{B}} = - ( N - 2 ) E_{AB} .   
\eeq
By using the Jacobi relation,
\beq
\label{A8}
\pa_{A} E_{BC} (-1)^{\varepsilon_{A} \varepsilon_{C}}  + cycle( A, B, C )  =
0,      
\eeq
one gets from (\ref{A7}),
\beq
\label{A9}
\pa_{A} V_{B} - \pa_{B} V_{A} (-1)^{\varepsilon_{A} \varepsilon_{B}}  =
E_{AB},    
\eeq
which is exactly the relation (\ref{i20.8}), where the $V_{A}$ is defined by just
(\ref{i20.7}),
\beq
\label{A10}
N^{C} = 2 V_{B}  E^{BC}.    
\eeq
Just when rewriting in (\ref{A7})  derivatives of $N^{C}$ in terms of derivatives of
$V_{B}$ coming from (\ref{A10}), there
appear the terms with derivatives of  $E_{CB}$ which cancel exactly the term
$(-N E_{AB})$ on the right-hand side in (\ref{A7}), due to the Jacobi relation  (\ref{A8}).

The general solution to Eq. (\ref{A9}) is given by
\beq
\label{A11}
V_{B} =
Z^{A}\bar{E}_{AB}+\pa_{B}F, \quad\bar{E}_{AB}=( Z^{C} \pa_{C} + 2)^{-1} E_{AB},    
\eeq
with $F $, $\varepsilon( F ) = 1$, being arbitrary Fermion.
With respect to the measure $\rho [ dZ ]$, the antisymmetry  of the $N$
requires
\beq
\div N = (-1)^{\varepsilon_{A}} \rho^{-1} \pa_{A}( \rho N^{A} )
= 0. 
\eeq

Here, we show in short that the general solution to Eq.
(\ref{A9}) has actually the form (\ref{A11}).
By multiplying the (\ref{A9}) by $Z^{A}$ from the left, we have
\beq
\label{A13}
( Z^{A} \pa_{A} + 1 ) V_{B} = Z^{A} E_{AB} + \pa_{B} ( Z^{A} V_{A} ). 
\eeq
Now, it is worthy to mention the two useful operator-valued
identities,
\beq
\label{A14}
( Z^{C} \pa_{C} + n )^{-1} Z^{A} = Z^{A} ( Z^{C} \pa_{C} + n + 1
)^{-1}, 
\eeq
\beq
\label{A15}
( Z^{C} \pa_{C} + n )^{-1} \pa_{A} = \pa_{A} ( Z^{C} \pa_{C} + n - 1)^{-1}. 
\eeq
Due to the latter identities, it follows immediately from (\ref{A13})
that the formula (\ref{A11}) holds with $F$ defined as
\beq
\label{A16}
F = Z^{A} ( Z^{C}\pa_{C} + 1 )^{-1} V_{A}.      
\eeq
If one inserts (\ref{A11}) into (\ref{A9}), the quantity (\ref{A16}) drops out, so
that (\ref{A9}) imposes no restrictions
on the Fermion F being thereby arbitrary. Thus, we have shown that
the (\ref{A11}) is just the general solution to (\ref{A9}).
By applying the operator $( Z^{D} \pa_{D} + 3 )^{-1}$ to (\ref{A8}) from the left,
and using (\ref{A15}), one gets the Jacobi identity, similar to (\ref{A8}),
as for $\bar{E}_{AB}$ in (\ref{A11}),
\beq
\label{A17}
\pa_{A} \bar{E}_{BC}(-1)^{ \varepsilon_{A} \varepsilon_{C} }+cycle( A,B,C )=0.   
\eeq
In turn, by using the latter identity, one can confirm, in an independent
way, that the solution (\ref{A11}) satisfies (\ref{A9}).

Next,  let us elucidate the formal essence of Eq. (\ref{A5}). By
inserting therein,
\beq
\label{A11.2}
N^{C} = N Z^{C},    
\eeq
and using then (\ref{A1}), (\ref{A2}) and (\ref{A4}), we have
on the left-hand side of (\ref{A5}),
\beq
\label{A12.2}
2 \Delta N^{C}  =  2 \Delta N Z^{C}  =  2 ( N + 2 ) \Delta Z^{C}  =  (N + 2
) E^{C},     
\eeq
which coincides exactly with the right-hand side in (\ref{A5}).  Thus, we have
confirmed again that the (\ref{A5}) is consistent with (\ref{A4}).

Finally, let us consider the relation
\beq
\nonumber
\label{A13.3}
&&N ( f, g ) = ( N f, g ) + ( f, N g )  -
f \overleftarrow{\pa}_{A} \big[N^{A}\overleftarrow{\pa_{C}} E^{CB} -\\
\nonumber
&&-( A \leftrightarrow B )
(-1)^{ ( \varepsilon_{A} + 1 ) ( \varepsilon_{B} + 1 ) } \big]
\overrightarrow{\pa}_{B} g  +
f \overleftarrow{\pa}_{A} ( N E^{AB} ) \overrightarrow{\pa}_{B} g =\\
\label{A13.3}
&&=( N f, g )  + (f, N g ) - 2 ( f, g ),    
\eeq
where we have used (\ref{A6}). The (\ref{A13.3}) tells us that the operator $( N - 2 )$
does differentiate the antibracket,
\beq
\label{A14.1}
( N - 2 ) ( f, g ) = ( ( N- 2 ) f, g ) + ( f, ( N - 2 ) g ),      
\eeq
which is the general-coordinate counterpart to (\ref{i26}).
\\

\appendix
\section*{Appendix B. Trivially-Deformed Extended Sigma-Model\footnote{In this Appendix,
we restrict ourselves with the use of the simplest power-counting
operator (\ref{i6.2}). }}
\setcounter{section}{2}
\renewcommand{\theequation}{\thesection.\arabic{equation}}
\setcounter{equation}{0}

Let \beq
\label{B.1}
\tau = \{ Z^{\alpha }\} = \{ Z^{A}, \ln t, \theta \},
\eeq
be the extended set of antisymplectic  variables we have
introduced in Secs. 2 and 3. Let us assume now that all the
variables (\ref{B.1}) are superfields depending on $2n$ Bosons
$u^{a}$ and $2n$ Fermions $\xi^{a}$, $a = 1, 2,...,2n$; these
variables are independent arguments of superfields (\ref{B.1}). Let
$D$ be the differential of De Rham, \beq \label{B.2} D = \xi^{a}
\pa_{a}, \quad \pa_{a} = \frac{\pa}{\pa u^{a}},
\quad   \varepsilon( D ) = 1, \quad D^{2} = 0.      
\eeq

The trivially-deformed extended sigma-model is defined by the action
\beq \label{B.3}
\Sigma = \int [du] [d\xi ] \mathcal{L},     
\eeq
with $\mathcal{L}$ being a Lagrange density,\footnote{In  the  case of  the
general  operator  (\ref{i15}),  the first term in  the
Lagrangian  (\ref{B.4}) should  be replaced  by:
$(1/2) \bar{Z}^{B} E_{BA} D \bar{Z}^{A} (-1)^{ \varepsilon_{A} }$,  with  the
zero  mode $\bar{Z}^{A}$ given by the formula (\ref{i7.5}).}
\beq \label{B.4}
\mathcal{L} =\frac{1}{2} t^{-2} Z^{B} E_{BA} D Z^{A}
(-1)^{\varepsilon_{A}} +
\frac{1}{2} ( \theta D \ln t + \ln t D \theta ) + T S,    
\eeq where $E^{AB}$ is a constant invertible antisymplectic  metric,
see (\ref{2.2}), and $E_{AB}$ is  the inverse to $E^{AB}$;  the
operator $T$ is defined in (\ref{i54})
\beq
T = 1 + \kappa \theta \Delta_{\tau *} ;     
\eeq
the Boson master action $S$ satisfies the  extended
trivially-deformed classical master equation
(\ref{i7.1})/(\ref{i7.2}),
\beq
( S, S )_{\tau *} = T^{-1} ( T S, T S )_{\tau} = 0.    
\eeq

One has to seek for a solution to that equation in the form similar
to (\ref{i5.21}), \beq \label{B.5} S = \sum_{k = -2}^{\infty} S_{ (k
|0) } t^{k} + \theta \sum_{k = 1}^{\infty}
S_{ (k|1) } t^{k} ,     
\eeq
where the component $S_{ (-2|0) }=\mathcal{S}$ is identified
with the classical nontrivially deformed proper action,
\beq
\label{B.6}
( \mathcal{S}, \mathcal{S} )_{*} = 0.     
\eeq

On the right-hand side in (\ref{B.4}), the kinetic part has the form
usual for sigma - models \cite{AKSZ,CF,BM4,BM5}, while the term $T
S$ is a natural counterpart to (\ref{5.27}). For the action
$\Sigma$, we have \beq \label{B.9} \frac{\delta}{ \delta Z^{B}}
\Sigma = t^{-2} E_{BA}\;t \!\;D \;t^{-1} Z^{A}
(-1)^{\varepsilon_{A}} + \pa_{B} ( T S ),    
\eeq \beq \label{B.10}
\frac{\delta}{\delta \ln t } \Sigma = D \theta -
t^{-2} Z^{B} E_{BA} D Z^{A} (-1)^{\varepsilon_{A}} +
\frac{\pa}{\pa {\ln t}} ( T S ),  
\eeq
\beq
\label{B.11}
\frac{\delta}{\delta \theta} \Sigma = D \ln t + \pa_{\theta} ( T S ).    
\eeq Thus, we get the following classical motion equations \beq
\label{B.12}
\nabla Z^{A} = 0, \quad \nabla \ln t = 0, \quad \nabla \theta = 0,    
\eeq
\beq
\label{B.13}
 \nabla = D + {\rm ad}_{\tau}( T S ),    
\eeq
where ${\rm ad}_{\tau}$ is the left adjoint of the $\tau$ extended
antibracket (\ref{i28}),
\beq \label{B.14}
{\rm ad}_{\tau}( X ) = ( X, ... )_{\tau}.   
\eeq

Now, let us define the functional extended antibracket,
\beq
&&[ F, G ]_{\tau} = \int [du] [d\xi]
F\; \frac{\overleftarrow{\delta}}{\delta Z^{\alpha}} ( Z^{\alpha}, Z^{\beta}
)_{\tau}\frac{\overrightarrow{\delta}}{\delta Z^{\beta}}\; G=\\
\nonumber
&&= \int [du] [d\xi] F \left[\frac{\overleftarrow{\delta}}{\delta Z^{A}} t^{2} E^{AB}
\frac{\overrightarrow{\delta}}{\delta Z^{B}} +
\frac{\overleftarrow{\delta}}{\delta Z^{A}} Z^{A}
\frac{\overrightarrow{\delta}}{\delta\theta} -
\frac{\overleftarrow{\delta}}{\delta \theta } Z^{A}
\frac{\overrightarrow{\delta}}{\delta Z^{A}} +
\frac{\overleftarrow{\delta}}{\delta \ln t}\frac{\overrightarrow{\delta}}{\delta \theta} -
\frac{\overleftarrow{\delta}}{\delta \theta}
\frac{\overrightarrow{\delta}}{\delta \ln t}\right] G,   
\eeq
where $Z^{\alpha}$ is the extended set (\ref{B.1}), and $F, G$ are functionals of
these variables. Then, we have for the action (\ref{B.3}),
that the following functional master equation holds,
\beq
\frac{1}{2} [ \Sigma, \Sigma ]_{\tau} = \int [du] [d\xi] \left( D \mathcal{L}
+ \frac{1}{2}(T S, T S )_{\tau} \right) = 0.    
\eeq

\vspace{4mm}

\appendix
\section*{Appendix C. Parametric Differential Equation for Star-Exponential}
\setcounter{section}{3}
\renewcommand{\theequation}{\thesection.\arabic{equation}}
\setcounter{equation}{0}

In Sec. 5,  we have noticed the formula (\ref{i62}) for the
star-exponential as derived in Ref. \cite{BB2}, in its App.
E. That derivation is very nice. However, here we would like to
re-derive the star exponential just from the first principle, by
resolving the basic parametric differential equation. What we mean
by the first principle is the definition
in (\ref{5.13}) of the star-product
\beq
\label{C.1}
(F * G) = T^{-1} ( (T F ) (T G) )  =
F G - \kappa \!\;\theta \!\;( F, G)_{\tau{*}} (-1)^{ \varepsilon(F)}.    
\eeq Besides , we will use the formula similar to (\ref{i63.2}) ($B$
is a Boson, $\varepsilon( B ) = 0$), \beq \label{C.2}
\exp_{*} \{ - B \} * ( B, \exp_{*} { B } )_{\tau{*}} = ( B, B)_{\tau{*}}. 
\eeq
So that
\beq
\label{C.3}
( B, \exp_{*}\{ B \} )_{\tau{*}} = ( B, B )_{\tau{*}} * \exp_{*}\{B\}. 
\eeq
Together with (\ref{C.1}),  (\ref{C.3}) enables us to present an alternative
derivation to the third equality in (\ref{i62}). The latter goes as follows. Let us
define ( $x$ is a Boson parameter, $\varepsilon( x ) = 0$),
\beq
\label{C.4}
U( x ) = \exp_{*}\{ x B \} ,   \quad    U( x = 0 ) = 1.  
\eeq
It follows from (\ref{C.1})  that
\beq
\nonumber
&&\pa_{x} U = B * U = B U - \kappa\;\theta\!\;( B, U )_{\tau{*}} =\\
\nonumber
&&=B U - x\!\; \kappa\;\theta\!\; ( B, B )_{\tau{*}} * U =\\
\label{C.5}
&&= (B - x\!\; \kappa\; \theta \!\;( B, B )_{\tau{*}} )\!\; U,    
\eeq where we have used  (\ref{C.1})  in the third equality. We have
also omitted the star, $*$ , just in  front of the rightmost $U$ in
the second line in (\ref{C.5}), because of the explicit presence of
the $\theta$ neighboring  to the left from $( B, B )_{\tau{*}}$,
see, again,  the second equality in (\ref{C.1}). The latter equality
in (\ref{C.5}) is just what we mean when saying about the basic
parametric differential equation. By integrating (\ref{C.5}), we get
\beq
\label{C.6}
U(x) = \exp\left\{ x B - \frac{1}{2} x^{2} \kappa\!
\;\theta \!\;( B, B )_{\tau{*}} \right\}.
\eeq By taking herein $x = 1$ , we obtain, finally
\beq \label{C.7}
\exp_{*} \{ B \} = \exp\left\{ B -
\frac{1}{2} \kappa\!\; \theta\!\; ( B, B )_{\tau{*}} \right\}, 
\eeq
which is  exactly the last equality in (\ref{i62}). The latter
generalizes to arbitrary star function $f_{*}(B)$ as follows
\beq
\label{C.8}
f_{*}(B)=
\exp\left\{-\frac{1}{2}\kappa\!\; \theta\!\; ( B, B )_{\tau{*}}\!\;
\frac{\pa^2}{\pa B^2}\right\}f(B),
\eeq
where,  given a regular function $f( B )$,  the corresponding  star function
$f_{*}( B )$  is defined similarly to the first and second equalities in
(\ref{i62}),
\beq
\label{C.9}
f_{*}( B ) =  T^{-1} f( T B ),    
\eeq
in terms of the operators (\ref{i54}) and (\ref{i55}).

If one introduces in (\ref{C.7}) a  polarization similar to (\ref{i32}),
\beq
\label{C.10}
B = m^{i} n_{i}, \quad \varepsilon( m^{i} ) = \varepsilon( n_{i} ) =
\varepsilon_{i},       
\eeq
then the formula (\ref{C.8}) generalizes to
\beq
\label{C.11}
f_{*}({\bf m})=  \exp\left\{  - \frac{1}{2} \kappa \!\;\theta\!\;
(-1)^{\varepsilon_{i}}
( m^{i}, m^{j} )_{\tau *} \frac{\overrightarrow{\pa}}{\pa m^{j}}
\frac{\overrightarrow{\pa}}{ \pa m^{i}}\right\} f({\bf m}),  
\eeq
where $f({\bf m})$ is a regular function of the components $m^{i}$,
and
\beq
\label{C.12}
f_{*}({\bf m})=  T^{-1}f(T{\bf m}).   
\eeq

\appendix
\section*{Appendix D. General Solution to Eq. (\ref{5.6})}
\setcounter{section}{4}
\renewcommand{\theequation}{\thesection.\arabic{equation}}
\setcounter{equation}{0}

Let us consider Eq. (\ref{5.6})  for the operator $T$,
\beq
\label{D.1}
[ \Delta_{\tau}, T ]  =  \kappa N_{\tau} \Delta_{\tau} T,\quad
[ N_{\tau}, T] = 0,     
\eeq
within the algebra
\beq
\label{D.2}
\theta^{2} = 0, \;\; \Delta^2_{\tau} = 0, \; \; [ \theta, \Delta_{\tau} ] =
N_{\tau},\; \;[ \theta, N_{\tau} ] = 0, \;\;  [ \Delta_{\tau}, N_{\tau} ] = 0.  
\eeq
Notice that the $T$- algebra spanned by the three basic elements
$\Delta_{\tau}, T, N_{\tau}$ is generated naturally via the
nilpotency condition,
$\Omega^{2} = 0$, imposed on the following Fermionic operator
\beq
\label{Omega}
\Omega = C \Delta_{\tau} + B T + A N_{\tau} + B C \kappa N_{\tau}
\Delta_{\tau} \bar{B},   
\eeq where $C$ is a Bosonic coordinate, while $B$ and $A$ are Fermonic
ones, $B^{2} = 0$, $A^{2} = 0$; $\bar{B}$ is a Fermionic canonical
momentum to  $B$,  $[ B, \bar{B} ] = 1$, $(\bar{B} )^{2} = 0$,
$(B{\bar B})^2=B{\bar B}$. Vice versa, let the $T$ - algebra holds. Then,
the nilpotency of  $\Omega$
(\ref{Omega}) can be easily  seen via rewriting
\beq
\Omega = C \tilde{\Delta}_{\tau}  +  B T+A N_{\tau},     
\eeq
where the operator
\beq
\tilde{\Delta}_{\tau} = \Delta_{\tau} ( 1 - B \kappa N_{\tau}\bar{B} ), 
\eeq
does satisfy
\beq
 \tilde{\Delta}_{\tau * ( \kappa B \bar{B} )} =
\tilde{\Delta}_{\tau} ( 1 - \kappa B \bar{B} N_{\tau} )^{-1}  =
\Delta_{\tau},     
\eeq
\beq
\tilde{\Delta}^2_{\tau} = 0, \quad [\tilde{\Delta}_{\tau}, B T] =0, 
\eeq
\beq
\label{D8}
U ^{-1} \Delta_{\tau} U = \tilde{\Delta}_{\tau},     
\eeq
\beq
\label{D.9}
U = \exp\{ - \kappa\!\; \theta \Delta_{\tau} B \bar{B} \Phi( Y ) \}
=  T^{-1}_{ ( \kappa B \bar{B} ) } = 1  -  \kappa B \bar{B} \theta \Delta_{\tau}. 
\eeq
\beq
\label{D.10}
U^{-1}  =  \exp\{ \kappa \theta \Delta_{\tau} B \bar{B} \Phi( Y ) \}  =
 T_{ ( \kappa B \bar{B} ) }   =
1  +  \kappa B \bar{B} \theta \Delta_{\tau * ( \kappa B \bar{B} )} =
1 +   \kappa B \bar{B} \theta \Delta _ {\tau_{*}}.      
\eeq
\beq
\label{D11}
\Phi(Y) = - Y^{-1} \ln( 1 - Y ),  \quad Y = \kappa N_{\tau}.   
\eeq
It is worthy to mention here that the exponential operator (\ref{D.10})
generalizes for arbitrary $\Phi( Y )$  to
\beq
\label{D.12}
\exp\{\kappa \theta \Delta_{\tau} B \bar{B} \Phi \}  =  1  +  \kappa\!\; \theta
\Delta_{\tau} B \bar {B} Y^{-1} ( \exp\{ Y \Phi \} - 1 ),    
\eeq
while the operator (\ref{D.9})  generalizes to (\ref{D.12}) with  $- \kappa$ standing for
$\kappa$ ( including the $\kappa$ entering the $Y$ ).
The choice (\ref{D11}) does satisfy
\beq
( -Y )^{-1} ( \exp\{ - Y \Phi \} - 1 ) = 1, \quad  Y^{-1} ( \exp\{ Y \Phi \} - 1 )
= (1 - Y )^{-1}.    
\eeq
It follows from (\ref{D8}) that the $U$-transformation (\ref{D.9}) and (\ref{D.10})
does Abelianize the operator $\Omega$ (\ref{Omega}), by eliminating
from the latter $T$ as well as $\bar {B}$,
\beq
U \Omega U^{-1} = \Omega_{\rm Abelian} =  C \Delta_{\tau} + B + A N_{\tau}. 
\eeq

General solution to  the equations (\ref{D.1}) can be
sought in the form
\beq
\label{D.3}
T=\theta \Delta_{\tau} A( N_{\tau},\Delta_{\tau})+B( N_{\tau},\Delta_{\tau} ).  
\eeq
It follows from the first in (\ref{D.1}) that
\beq
\label{D.4}
N_{\tau} \Delta_{\tau} \big( ( 1 - \kappa N_{\tau}) A( N_{\tau},\Delta_{\tau}) -
\kappa B(N_{\tau},\Delta_{\tau} ) \big)  =  0.    
\eeq
In turn, it follows from (\ref{D.4}) that
\beq
\label{D.5}
A( N_{\tau},\Delta_{\tau} ) = \kappa ( 1 - \kappa N_{\tau} )^{-1} B( N_{\tau},
\Delta_{\tau} ) ,   
\eeq
where we  have included into $B$ the zero mode of the operator $N_{\tau}
\Delta_{\tau}$.
By inserting (\ref{D.5}) into (\ref{D.3}),
we get
\beq
\label{D.6}
T = \left ( 1 + \kappa \theta \Delta_{\tau} ( 1 - \kappa N_{\tau} )^{-1}\right)
B(N_{\tau}, \Delta_{\tau}).      
\eeq
Here, the overall factor $B$ remains arbitrary; that is a natural arbitrariness
in the general solution for $T$.   On the other hand, as the operator $T$ should
be equal to $1$ at $\kappa = 0$, it follows that
we should choose $B = 1$. Thus, we arrive at the formula (\ref{i54}).

Finally, let us consider the dual to Eq.(\ref{D.1}), for the inverse
$T^{-1}$,
\beq
\label{D.7}
[ \Delta_{\tau}, T^{-1} ] =  - T^{-1} \kappa N_{\tau} \Delta_{\tau},\quad
[N_{\tau}, T^{-1} ] = 0. 
\eeq
In principle, we could analyze these equations in the same way as we have
done as to the equations (\ref{D.1}). However, it is much simpler to consider directly
the inverse to the general solution (\ref{D.6}),
\beq
\label{D.8}
T^{-1} = ( B( N_{\tau}, \Delta_{\tau} ) )^{-1}
( 1 - \kappa\!\; \theta\!\;\Delta_{\tau} ).      
\eeq
For the reason mentioned just below (\ref{D.6}), we have to choose $B = 1$, so that
we arrive at the formula (\ref{i55}).
\\

\end{document}